\documentclass[aps,pra,twocolumn,superscriptaddress,showpacs,showkeys,amsmath,amssymb]{revtex4}

\usepackage{amsfonts}
\usepackage{amssymb,amsmath}
\usepackage{mathrsfs}
\usepackage{latexsym}
\usepackage{amsmath}
\usepackage[cp1251]{inputenc}
\usepackage{graphicx}
\usepackage{dcolumn}
\usepackage{bm}
\usepackage{color}

\RequirePackage{ifthen}
\RequirePackage[pdfstartview=FitH]{hyperref}
\begin{document}
	
\title{Evolution of a single spin in ideal Bose gas at finite temperatures}
\author{O.~Hryhorchak}\affiliation{Professor Ivan Vakarchuk Department for Theoretical Physics, Ivan Franko National University of Lviv, 12 Drahomanov Street, Lviv, Ukraine}
\author{G.~Panochko} \affiliation{Department of Optoelectronics and Information Technologies, Ivan Franko National University of Lviv, 107 Tarnavskyj Str., Lviv, Ukraine}
\author{V.~Pastukhov\footnote{e-mail: volodyapastukhov@gmail.com}}\affiliation{Professor Ivan Vakarchuk Department for Theoretical Physics, Ivan Franko National University of Lviv, 12 Drahomanov Street, Lviv, Ukraine}

\date{\today}

	
	
\begin{abstract}
	We study the finite-temperature dynamics of non-interacting bosons with a single static spinful impurity immersed. A non-zero contact boson-impurity pairwise interaction is assumed only for the spin-up impurity state. By tracing out bosonic degrees of freedom, the exact time evolution of the impurity spin is calculated for pure and mixed initial ensembles of states. The time-dependent momentum distribution of bosons initially created in the Bose-condensed state and driven by the interaction with spin is analyzed.
\end{abstract}
	
\maketitle
\section{Introduction}
A static (infinite-mass) impurity in various media is a paradigmatic model for understanding few- and many-body physics. Being a good playground for testing various approximate theories, this model provides the first, typically an exactly solvable, insight into the system's properties. The presence of a spin degree of freedom for the impurity immersed in bosonic baths allows for a number of practical applications, including non-destructive quantum probing \cite{Hangleiter_2015,Elliott_2016,Bouton_2020}, ultra-low-temperature thermometry \cite{Jevtic_2015,Yuan_2023}, etc.

Here, we consider and exactly calculate the dynamics of the spinful impurity interacting with the ideal gas of spinless particles. Previous studies were mainly concerned with the weakly-interacting Bose gas and treated the impurity with the approximate Fr\"ohlich Hamiltonian. Particularly, the relation between the spin superposition state dephasing and the fluctuations of BEC's phase field was established in \cite{Bruderer_2006}. Detailed characterization of dephasing and the decoherence dynamics for single atoms trapped in an optical lattice and loaded in weakly-interacting BEC were investigated in \cite{Klein_2007}. The simplified treatment \cite{Ng_2008} of spinful impurity immersed in the one- and two-component BECs, which restricts consideration to a single-mode approximation for bosons, proposed the decoherence probe by measuring the spin state. Experimental observation of the individual impurity spin dynamics in these systems was reported in \cite{Schmidt_2018}.

Spinless-impurity decoherence in the dilute Bose gas has been extensively analyzed both theoretically \cite{Nielsen_2019} and experimentally \cite{Skou_2021,Adam_2022}, mostly in relation to Bose polarons. Very promising in the context of impurities strongly coupled to various media are variational methods \cite{Liu_2019,Dzsotjan_2020,Pena_Ardila_2021}. The real-time dynamics of the finite-mass impurity after switching on interaction with the ideal Bose gas cannot be found exactly in any spatial dimension and requires some simplified consideration, for instance, the $t$-matrix approximation \cite{Volosniev_2015}. The exact analysis \cite{Drescher_2021} of the interaction quench is possible only in the case of a static impurity.

To some extent, our model is identical to the Rydberg polaron in an ideal Bose gas. The absorption spectrum of this system was obtained theoretically in \cite{Schmidt_2016}. A later experiment \cite{Camargo_2018} reported the creation of Rydberg Bose polarons and confirmed theoretical predictions to a remarkable degree. Recent work \cite{Durst_2024} outlined three distinct regimes of the Rydberg impurity as the density of bosons increases. Another possible application \cite{Lee_2023} of the presented model is the Bardeen--Cooper--Schrieffer superconductor with magnetic impurities in a magnetic field.
	
An important conclusion of the last decade of the Bose-polaron studies (see \cite{Mistakidis_2023,Grusdt_2025} for review) is that the Fr\"ohlich model is not fully adequate for these systems in the strongly-interacting regime. Even in the Bogoliubov approximation, the role of the two-phonon processes becomes defining when boson-impurity coupling increases. The latter circumstance should profoundly affect the spin dynamics and decoherence properties. Even though the environment of interacting bosons is more realistic from an experimental point of view, a complete theoretical treatment of spin evolution in such a medium is challenging, and the present analysis is the first step toward addressing this problem.

\section{Formulation}
\subsection{Model description}
The consider model describes a system of macroscopic number $N$ of mutually non-interacting Bose atoms loaded in a large volume $V$ with periodic boundary conditions that interact, through the (regularized) $\delta$-potential, with a (pseudo) spin-$\frac{1}{2}$ static (infinitely-heavy) particle (impurity). It should not necessarily be a physical spin but rather any two-state degree of freedom. Assume that the magnitude of the coupling constant depends on the spin state of the impurity, and even more, atomic collisions can provoke the spin overturn, i.e., the system's Hamiltonian is
\begin{eqnarray}\label{H}
	H=\sum_{\bf k}\varepsilon_kb^{\dagger}_{\bf k}b_{\bf k}
	+\frac{1}{V}\sum_{{\bf k},{\bf q}}b^{\dagger}_{\bf k}b_{\bf q}\sum_{\sigma, \sigma'}g_{\sigma \sigma'}|\sigma\rangle\langle \sigma'|,
\end{eqnarray}
where $|\sigma\rangle\langle \sigma'|$ is the projection operator acting on spin variables and bosonic creation (annihilation) operators $b^{\dagger}_{\bf k}$ ($b_{\bf k}$) satisfy canonical commutation relations $[b_{\bf k},b^{\dagger}_{{\bf k}'}]=\delta_{{\bf k},{\bf k}'}$ (and mutually commuting all others pairs). The dispersion relation of atoms is Galilean-invariant $\varepsilon_{\bf k}=\frac{{\bf k}^2}{2m}$ ($\hbar=1$) and for the Hamiltonian to be hermitian matrix, couplings $g_{\uparrow \uparrow}$ and $g_{\downarrow\downarrow}$ should be real, while $g^*_{\downarrow\uparrow}=g_{\uparrow\downarrow}$. Further analysis can be easily extended to an arbitrary number of static spinful impurities located in coordinates $\{{\bf r}_j\}$. In this case, the interaction term in (\ref{H}) should read 
\begin{eqnarray}\label{Phi}
\frac{1}{V}\sum_{{\bf k},{\bf q}}b^{\dagger}_{\bf k}b_{\bf q}\sum_{j}e^{-i({\bf k}-{\bf q}){\bf r}_j}\sum_{\sigma, \sigma'}g_{\sigma \sigma'}|\sigma\rangle_j \,_j\langle \sigma'|.
\end{eqnarray}
The latter model is interesting from the perspective of the collective magnetic properties of the system with a macroscopic number of spins. The bosonic medium induces the effective two-body, three-body, and higher-order inter-spin interactions. However, all of them are of the Ising type. This is seen from the following arguments. There is a unitary transformation that diagonalizes the spin part of the Hamiltonian (\ref{H}) with potential energy (\ref{Phi}). This transformation does not affect the partition function and, consequently, the thermodynamics of the system. The same trick can be utilized for our calculations of the time evolution problem, i.e., start from the diagonal ($g_{\uparrow\downarrow}=0$), in the spin sector, Hamiltonian (\ref{H}) \cite{Shashi_2014} and study the evolution of an arbitrary initial spin state. Since $g_{\uparrow \uparrow}|\uparrow\rangle\langle \uparrow|+g_{\downarrow \downarrow}|\downarrow\rangle\langle \downarrow|=g_{\downarrow \downarrow}+(g_{\uparrow \uparrow}-g_{\downarrow \downarrow})|\uparrow\rangle\langle \uparrow|$ (here and below unit matrix is not explicitly written down), and we are only interested in the non-trivial spin dynamics, we put $g_{\downarrow\downarrow}=0$ and are left with a single non-zero coupling $g_{\uparrow \uparrow}=g$.

A standard idea for treating Hamiltonian (\ref{H}) in the BEC phase is the separation of the condensate mode $b^{\dagger}_{\bf 0},b_{\bf 0}\to \sqrt{N_{\bf 0}}$, with the further neglecting of the interaction between non-condensed bosons
\begin{eqnarray}\label{Phi_simple}
	\frac{\sqrt{N_{\bf 0}}}{V}\sum_{{\bf k}\neq {\bf 0}}\left(b^{\dagger}_{\bf k}+b_{-{\bf k}}\right)\sum_{\sigma, \sigma'}g_{\sigma \sigma'}|\sigma\rangle\langle \sigma'|.
\end{eqnarray}
The resulting Hamiltonian describes the so-called dephasing model. By the further inclusion of the transverse magnetic field acting on the spin degrees of freedom, which generates non-trivial dynamics, one complicates the Hamiltonian to the multimode variation \cite{Seke_1985} of the celebrated Jaynes-Cummings model (see \cite{Shore_1993,Larson_2021} for review). However, below we show that the replacement of the original interaction potential by the simplified one (\ref{Phi_simple}) is incomplete and incorrectly captures properties of the system even at absolute zero. The finite-temperature region is characterized \cite{Lausch_2018} by an increasing impact from collisions of non-Bose-condensed atoms, which are neglected in (\ref{Phi_simple}).

\subsection{Time evolution}
The protocol suggests that initially non-interacting ($g=0$) subsystems -- a spin and a Bose gas -- are prepared in pure or mixed states, and then undergo, at $t=0$, a sudden interaction switch-on from $g=0$ to $g\neq 0$. The dynamics of the system in the pure state is governed by the formal solution of the Schr\"odinger equation
\begin{eqnarray}\label{}
&&|\Psi(t)\rangle=e^{-itH} |\Psi(0)\rangle,\label{Psi} \\ &&|\Psi(0)\rangle=|\{N_{\bf k}\}\rangle\otimes|\psi\rangle\label{Psi_0},
\end{eqnarray}
where the eigenstate of free bosons reads $|\{N_{\bf k}\}\rangle=\prod_{\bf k}\frac{(b^{\dagger}_{\bf k})^{N_{\bf k}}}{\sqrt{N_{\bf k}!}}|0\rangle$, with $|\psi\rangle$ being an arbitrary spin state and $|0\rangle$ is a normalized bosonic vacuum. Note that the constraint $\sum_{\bf k}N_{\bf k}=N$ should be applied, and any initial state can be represented as a linear combination of $|\Psi(0)\rangle$s. From (\ref{Psi}) and (\ref{Psi_0}), it is clear that one must calculate operator
\begin{eqnarray}\label{b_t}
b^{\dagger}_{\bf k}(t)=e^{-itH}b^{\dagger}_{\bf k}e^{itH},
\end{eqnarray}
in order to reveal the system's dynamics. By using explicit representation
\begin{eqnarray}\label{}
b^{\dagger}_{\bf k}(t)=\sum_{\bf q}C_{{\bf q},{\bf k}}(t)b^{\dagger}_{\bf q},
\end{eqnarray}
and `equation of motion' $i\partial_tb^{\dagger}_{\bf k}(t)=[H,b^{\dagger}_{\bf k}(t)]$ for the operator (\ref{b_t}), we can write down the system of coupled differential equations
\begin{eqnarray}\label{C_Eq}
i\partial_tC_{{\bf q},{\bf k}}(t)=\varepsilon_{\bf q}C_{{\bf q},{\bf k}}(t)+ \frac{1}{V}\sum_{\bf p}C_{{\bf p},{\bf k}}(t)g|\uparrow\rangle\langle \uparrow|,
\end{eqnarray}
for unknown coefficients $C_{{\bf q},{\bf k}}(t)$ (actually, square matrices acting in the Hilbert space of spin variable). The commutation relations between equal-time operators $[b_{\bf k}(t),b^{\dagger}_{{\bf k}'}(t)]=\delta_{{\bf k},{\bf k}'}$ set the unitarity condition $\sum_{\bf q}C^{\dagger}_{{\bf q},{\bf k}}(t)C_{{\bf q},{\bf k}'}(t)=\delta_{{\bf k},{\bf k}'}$ on these matrices. With the initial condition $C_{{\bf q},{\bf k}}(0)=\delta_{{\bf q},{\bf k}}$ taken into account, a formal solution to Eq.~(\ref{C_Eq}) reads
\begin{eqnarray}\label{C_sol}
&&C_{{\bf q},{\bf k}}(t)=\delta_{{\bf q},{\bf k}}e^{-it\varepsilon_{\bf q}}\nonumber\\
&&-\frac{i}{V}\sum_{\bf p}\int^{t}_0dt'e^{-i(t-t')\varepsilon_{\bf q}} C_{{\bf p},{\bf k}}(t')g|\uparrow\rangle\langle \uparrow|.
\end{eqnarray}
A substitution of an arbitrary interaction matrix (instead of $g|\uparrow\rangle\langle \uparrow|$) in the above formula extends our consideration to the most generic case. Passing to the Laplace transform $\Gamma_{{\bf q},{\bf k}}(\omega)=\int_{0}^{\infty}dte^{-t\omega}C_{{\bf q},{\bf k}}(t)$, Eq.~(\ref{C_sol}) can be converted into the linear algebraic one
\begin{eqnarray}\label{}
(\omega+i\varepsilon_{\bf q})\Gamma_{{\bf q},{\bf k}}(\omega)=\delta_{{\bf q},{\bf k}}
-\frac{i}{V}\sum_{\bf p}\Gamma_{{\bf p},{\bf k}}(\omega)g|\uparrow\rangle\langle \uparrow|.
\end{eqnarray}
Iterating the latter equation allows us to guess the structure of its solution
\begin{eqnarray}\label{Gamma_omega}
\Gamma_{{\bf q},{\bf k}}(\omega)=\frac{\delta_{{\bf q},{\bf k}}}{\omega+i\varepsilon_{\bf q}}
-\frac{i}{V}\frac{\mathcal{T}(\omega)|\uparrow\rangle\langle \uparrow|}{(\omega+i\varepsilon_{\bf q})(\omega+i\varepsilon_{\bf k})},
\end{eqnarray}
where the $T$-matrix of a single boson in the presence of an impurity is
\begin{eqnarray}
\mathcal{T}^{-1}(\omega)=g^{-1}+\frac{1}{V}\sum_{\bf p}\frac{1}{\varepsilon_{\bf p}-i\omega}.
\end{eqnarray}
With the known Laplace transform (\ref{Gamma_omega}), one reproduces function $C_{{\bf q},{\bf k}}(t)=\int_{\delta-i\infty}^{\delta+i\infty}\frac{d\omega}{2\pi i}e^{\omega t}\Gamma_{{\bf q},{\bf k}}(\omega)$ via the Mellin transform. The sum in definition of the $T$-matrix can be calculated in the thermodynamic limit to yield $\mathcal{T}^{-1}(\omega)=g^{-1}\left[1-a\sqrt{2m(-i\omega)}\right]$ (here $a$ is the $s$-wave scattering length and the renormalized coupling $g=\frac{2\pi a}{m}$). In this case, the inverse Laplace transform is computed by contour integration in the complex plane, yielding a result involving non-elementary functions (see Appendix).

\subsection{Spin reduced density matrix}
Having identified the system's wave function at every time moment $|\Psi(t)\rangle=\prod_{\bf k}\frac{[b^{\dagger}_{\bf k}(t)]^{N_{\bf k}}}{\sqrt{N_{\bf k}!}}|0\rangle\otimes|\psi\rangle$, we are in position to obtain the reduced density matrix of a spin. Indeed, by tracing out the bosonic degrees of freedom in the density matrix of the full system described by the pure ensemble of states, one computes the reduced density matrix of the spin subsystem
\begin{eqnarray}\label{rho_def}
\rho(t)=\textrm{Sp}_b\left\{|\Psi(t)\rangle\langle \Psi(t)|\right\}.
\end{eqnarray}
Needless to say, $\rho(t)$ contains all information about the properties of spin in the bosonic medium. The calculation of trace in (\ref{rho_def}) is straightforward in the second quantization representation. Alternatively, one may use the completeness of coherent states $|\beta\rangle=\prod_{\bf k}e^{-|\beta_{\bf k}|^2/2+\beta_{\bf k}b^{\dagger}_{\bf k}}|0\rangle$ to obtain $\rho(t)$. The result can be most simply outlined if we explicitly specify the initial spin state $|\psi\rangle=A|\uparrow\rangle+B|\downarrow\rangle$ (with normalization $|A|^2+|B|^2=1$ and either $A$ or $B$ chosen real positive) and the diagonal matrix element (the only non-zero) of $C_{{\bf q},{\bf k}}(t)$ as $C^{\uparrow\uparrow}_{{\bf q},{\bf k}}(t)$ and $C^{\downarrow\downarrow}_{{\bf q},{\bf k}}(t)$. The final answer is
\begin{eqnarray}
\rho(t)=|A|^2|\uparrow\rangle\langle \uparrow|+A^*Bf^*(t)|\downarrow\rangle\langle \uparrow|\label{rho}\nonumber\\
+AB^*f(t)|\uparrow\rangle\langle \downarrow|+|B|^2|\downarrow\rangle\langle \downarrow|,\\
f(t)=\prod_{\bf k}\left\{\sum_{\bf q}C^{\uparrow\uparrow}_{{\bf q},{\bf k}}(t)\left[C^{\downarrow\downarrow}_{{\bf q},{\bf k}}(t)\right]^*\right\}^{N_{\bf k}}.
\end{eqnarray}
Taking into account explicit expression for $C_{{\bf q},{\bf k}}(t)$ from the Appendix, the introduced function $f(t)$ can be written as an exponent in the thermodynamic limit.

The above calculation scheme straightforwardly extends to a case of a mixed ensemble. Indeed, any density matrix of the initial mixed state can be expanded in the basis $|\{N_{\bf k}\}\rangle\otimes|\sigma\rangle$ with the subsequent computation of the time-evolution. A natural choice for an initial state is given by a tensor product of arbitrary density matrix $\rho_s$ of spin and the density matrix of bosons $\frac{1}{Z}\sum_{\{N_{\bf k}\}}\delta_{\sum_{\bf k}N_{\bf k},N}e^{-\sum_{\bf k}\varepsilon_{\bf k}N_{\bf k}/T}|\{N_{\bf k}\}\rangle\langle \{N_{\bf k}\}|$ (with $ Z=\sum_{\{N_{\bf k}\}}\delta_{\sum_{\bf k}N_{\bf k},N}e^{-\sum_{\bf k}\varepsilon_{\bf k}N_{\bf k}/T}$) in thermal equilibrium. Rewriting Kronecker delta via integral $\delta_{\sum_{\bf k}N_{\bf k},N}=\int_{0}^{2\pi}\frac{d\phi}{2\pi}e^{i\phi(\sum_{\bf k}N_{\bf k}-N)}$, making use of the time dependence for $|\{N_{\bf k}\}\rangle$, computing the trace by performing summations over $\{N_{\bf k}\}$ and then calculating the remaining integral by means of the steepest-descent method, we obtain
\begin{eqnarray}\label{rho_T}
\rho_T(t)=\rho_{\uparrow\uparrow}|\uparrow\rangle\langle \uparrow|+\rho_{\downarrow\uparrow}f^*_T(t)|\downarrow\rangle\langle \uparrow|\nonumber\\
+\rho_{\uparrow\downarrow}f_T(t)|\uparrow\rangle\langle \downarrow|+\rho_{\downarrow\downarrow}|\downarrow\rangle\langle \downarrow|,
\end{eqnarray}
where $\rho_{\sigma\sigma'}$ are the elements of initial matrix $\rho_s$, and the temperature-dependent function
\begin{eqnarray}\label{f_T}
\ln f_T(t)=\sum_{\bf k}\langle N_{\bf k}\rangle\left\{\sum_{\bf q}C^{\uparrow\uparrow}_{{\bf q},{\bf k}}(t)\left[C^{\downarrow\downarrow}_{{\bf q},{\bf k}}(t)\right]^*-1\right\}.
\end{eqnarray}
Note that the sum over ${\bf k}$ in the function $\ln f_T(t)$ disappears at absolute zero, where all bosons are in a state with minimal energy, i.e. $\langle N_{\bf k}\rangle|_{T=0}=N\delta_{{\bf k}, {\bf 0}}$.

There are a few notes about the obtained formulas. First, they are asymptotically valid in the thermodynamic limit. The Bose-Einstein distribution $\langle N_{\bf k}\rangle=\left[e^{(\varepsilon_{\bf k}-\mu)/T}-1\right]^{-1}$ contains the chemical potential $\mu$ of an ideal Bose gas, which appeared as a generally complex-valued \cite{Schelle_2017}
saddle point in the integration over $\phi$ and is determined by equation $\sum_{\bf k}\langle N_{\bf k}\rangle=N$. In the normal phase at thermodynamic equilibrium, the chemical potential is real and negative $\mu<0$, producing a gap in the excitation spectrum effectively. Below the critical point ($T<T_0$) where $\mu=0$, the modified equation $\langle N_{\bf 0}\rangle+\sum_{{\bf k}\neq {\bf 0}}\langle N_{\bf k}\rangle=N$ allows the calculation of Bose-Einstein condensate $\langle N_{\bf 0}\rangle$. In principle, both equations for $\rho(t)$ and $\rho_T(t)$ remain true for an arbitrary interaction matrix in the Hamiltonian (\ref{H}). Secondly, while computing Eq.~(\ref{f_T}), we implicitly assumed that the presence of a single spin cannot drastically change the properties of the macroscopic system. This is not always the case for mutually non-interacting bosons, where the bound states in the single-particle spectrum can cause the collapse \cite{Panochko_2021} of the macroscopic system. To cure this non-thermodynamic behavior, one can provide weak repulsion between Bose atoms or (and) the strong confining potential \cite{Hryhorchak_2023} in one or two spatial dimensions.

\section{Results and discussion}
\begin{figure}[h!]
	{\includegraphics
		[width=0.4
		\textwidth,clip,angle=-0]{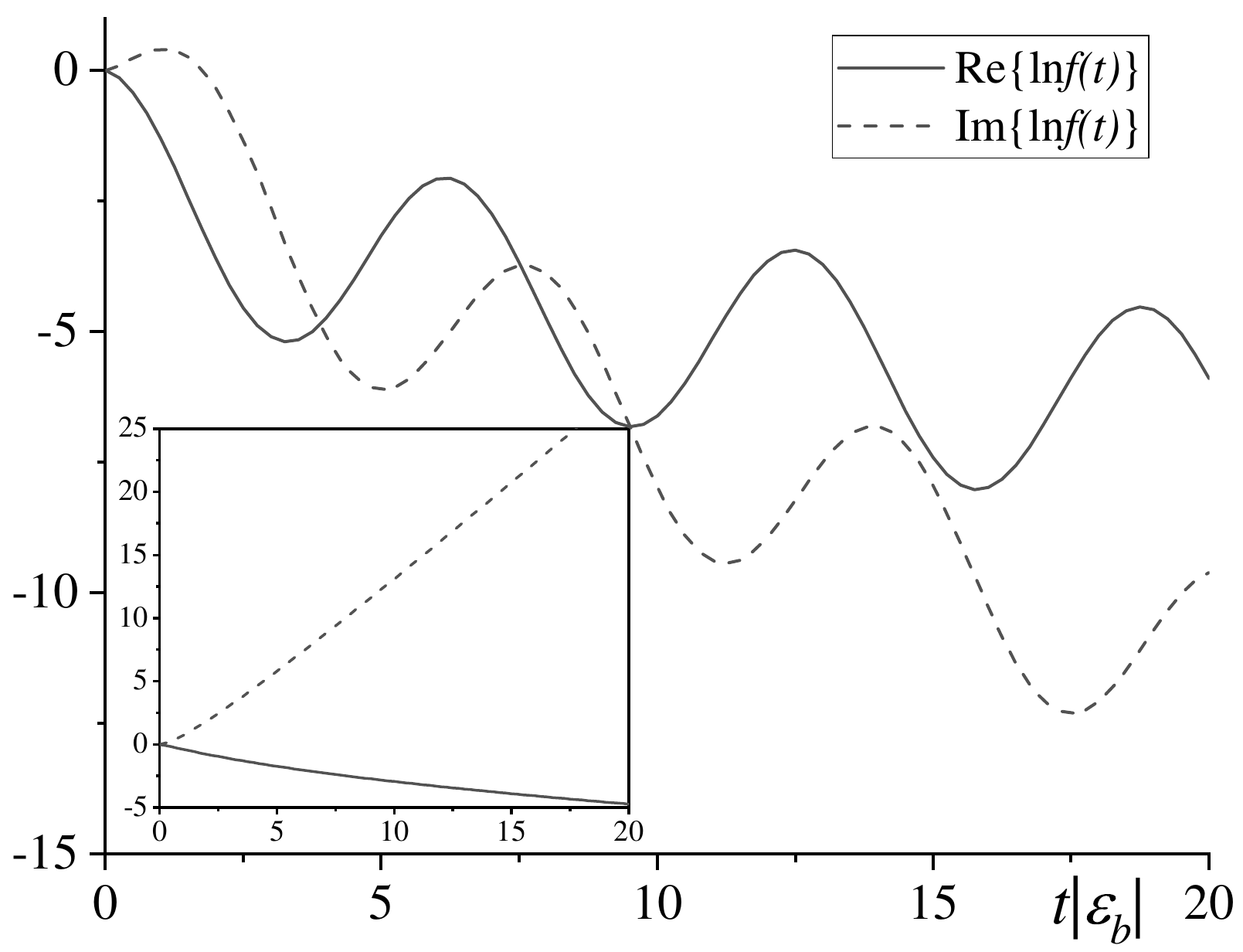}}
	\caption{Real (solid line) and imaginary (dashed line) parts of $\ln f(t)$ as a function of dimensionless time variable $t|\epsilon_b|$ at $n|g|/|\epsilon_b|=1$. The main panel shows results for $a>0$; the inset shows results for $a<0$.}\label{lnf(t)_fig}
\end{figure}

\begin{figure}[h!]
	{\includegraphics
		[width=0.4
		\textwidth,clip,angle=-0]{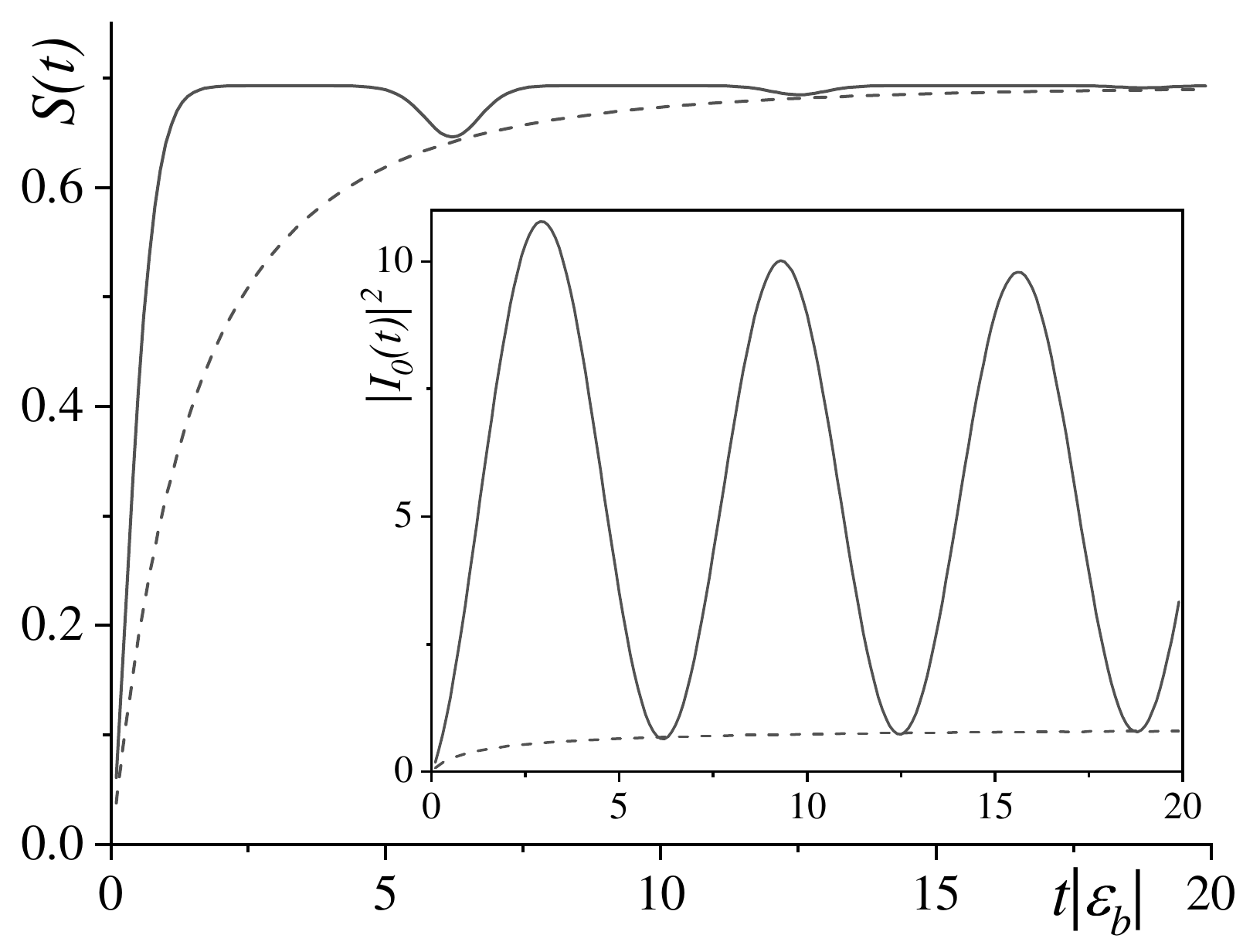}}
	\caption{The time dependence of the entanglement entropy (main panel) and dimensionless Tan's contact $|I_{\bf 0}(t)|^2$ (inset) for $a>0$ (solid lines) and $a<0$ (dashed lines). The coupling is $n|g|/|\epsilon_b|=1$ in both cases, and the initial spin state is fixed by $|A|^2=|B|^2=1/2$. }\label{S(t)_fig}
\end{figure}

It is natural to start by discussing the results obtained by first considering an initial state in which all bosons occupy the zero-momentum energy level. This is the ground state of an ideal Bose gas without the impurity immersed. The conserved, during the unitary evolution, energy of the system with an arbitrarily chosen spin state $|\Psi(0)\rangle=\frac{(b^{\dagger}_{\bf 0})^{N}}{\sqrt{N!}}|0\rangle\otimes|\psi\rangle$ is zero (due to peculiarities of the $\delta$-like interaction in three dimensions). Within this initial state, the function $f(t)$ in the limit $N\to \infty$, $V\to \infty$ (but with the ratio $n=N/V$ being finite) reads
\begin{eqnarray}
\ln f(t)=n|g|\left.\frac{\partial I_{\bf k}(t)}{\partial \varepsilon_{\bf k}}\right|_{{\bf k}= {\bf 0}},
\end{eqnarray}
where $I_{\bf k}(t)$ is different for positive and negative $g$s and given explicitly in Appendix. The imaginary part of $\ln f(t)$ is responsible for the oscillations of the off-diagonal elements of the spin reduced density matrix. In contrast, the non-positive real part provides the decoherence of a spin in the bosonic environment. The result for $\ln f(t)$ can represented as a density-dependent factor $n|g|/|\epsilon_b|$, where $\epsilon_b=-\frac{1}{2ma^2}$ is the bound-state energy (for $g>0$) of a single boson in the presence of a spin-up impurity, multiplied by a universal function of dimensionless time $\tilde{t}=t|\epsilon_b|$. The appropriate curves are presented in Fig.~\ref{lnf(t)_fig}.
Periodic behavior of the real part $\Re \ln f(t)$ indicates partial spin recoherence. We have also calculated numerically (see Fig.~\ref{S(t)_fig})
the time dependence of the entanglement entropy
\begin{eqnarray}
S(t)=-\textrm{Sp}_s\left\{\rho(t)\ln\rho(t)\right\},
\end{eqnarray}	
for a spin in the bosonic bath. Since, at late times, the modulus of the function $f(t)$ tends to zero exponentially fast
\begin{eqnarray}
|f(t)|_{t\to \infty}\sim e^{-\frac{n|g|}{|\epsilon_b|}\left[\sqrt{\frac{2\tilde{t}}{\pi}}+2\theta(a)(1-\cos \tilde{t})+1\right]},
\end{eqnarray}
the behavior of the density matrix eigenvalues is $|A|^2$ and $1-|A|^2$, respectively; therefore, the maximum of the entanglement entropy is reached at $S(\infty)=-|A|^2\ln |A|^2-(1-|A|^2)\ln(1-|A|^2)$. The calculated behavior of the entanglement entropy looks like the average entropy production obtained in \cite{Islam_2021} for a two-level system interacting with an electromagnetic field. Technically, bosons in the BEC state are similar to the radiation field at thermal equilibrium. 

\begin{figure}[h!]
		{\includegraphics
			[width=0.4
		\textwidth,clip,angle=-0]{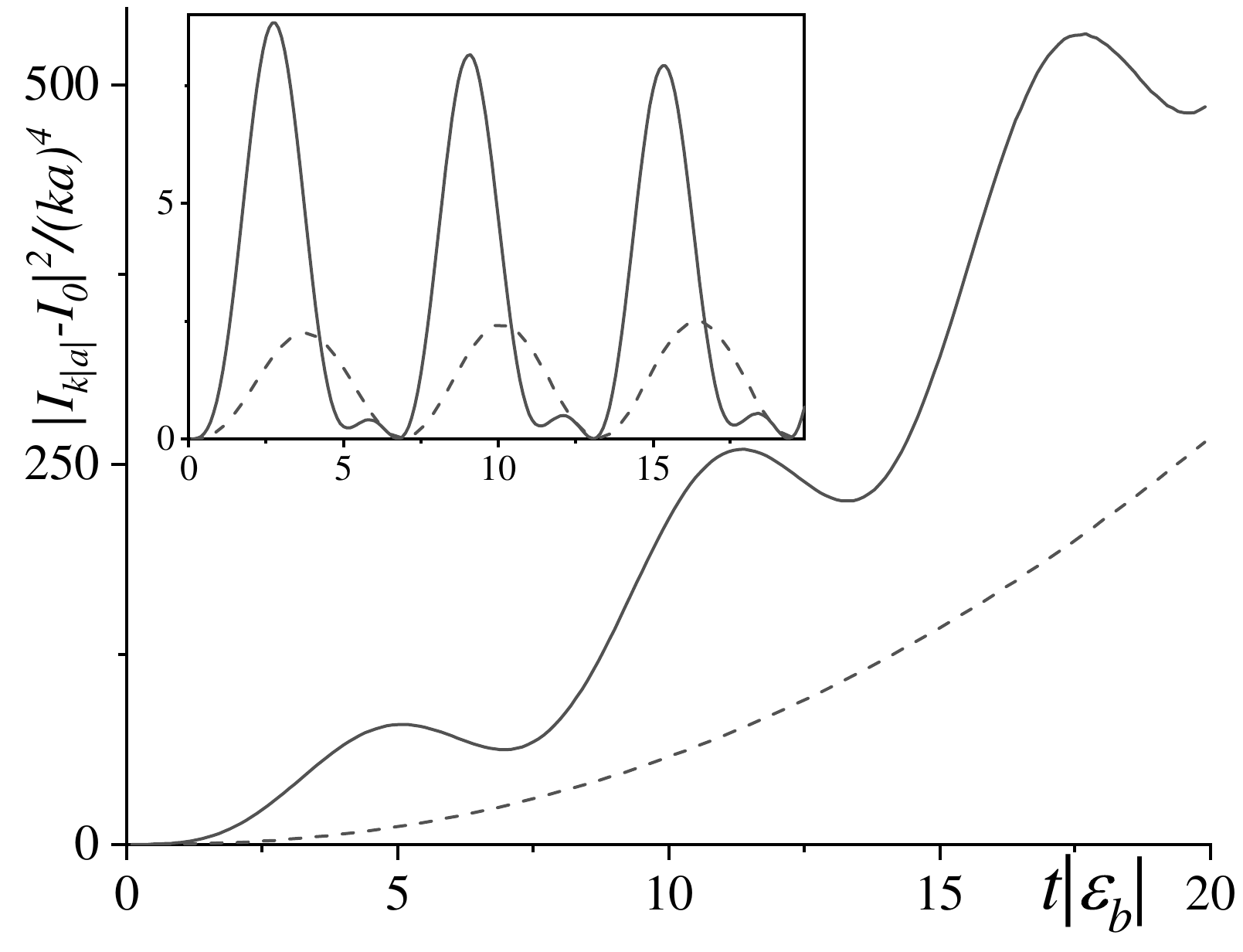}}
		\caption{Dimensionless fraction determining time dependence of momentum distribution for $|{\bf k}a|=0.1$ (main panel) and $|{\bf k}a|=1.0$ (inset). Solid and dashed lines refer to repulsive ($a>0$) and attractive ($a<0$) interactions, respectively.}\label{moment_dis_fig}
\end{figure}

Interaction with spin affects the momentum distribution of the initially Bose-Einstein-condensed bosons. The time dependence of the average number of atoms with momentum ${\bf k}$ is easily computable
\begin{eqnarray}\label{av_N_k}
&&\langle \Psi(t)|N_{\bf k}|\Psi(t)\rangle=N\delta_{{\bf k},{\bf 0}}\nonumber\\
&&+N|A|^2\left[|C^{\uparrow\uparrow}_{{\bf 0},{\bf k}}(t)|^2-\delta_{{\bf k},{\bf 0}}\right].
\end{eqnarray}
Here, the second term is of order $1/V$ and consequently disappears in the thermodynamic limit (on the other hand, by summing both sides over ${\bf k}$ with taking into account the unitarity condition on $C_{{\bf q},{\bf k}}(t)$, the boson number conservation is demonstrated). At this point, however, one can speculate considering not a single, but rather a macroscopic number $\mathcal{N}$ (such that $\mathcal{N}/V$ is fixed when $V\to \infty$) of mutually distant spins. The latter requirement is mandatory to avoid the impact of the induced spin-spin interaction. Assuming every spin is in state $|\psi\rangle_j$ (with coefficients $A_j$ and $B_j$), the only modification of Eq.~(\ref{av_N_k}) required is the replacement $|A|^2\to \sum_{j}|A_j|^2$. Making use of an explicit formula for $C^{\uparrow\uparrow}_{{\bf 0},{\bf k}}(t)$, one gets
\begin{eqnarray}\label{N_k}
&&\langle \Psi(t)|N_{\bf k}|\Psi(t)\rangle=\langle \Psi(t)|N_{\bf 0}|\Psi(t)\rangle\delta_{{\bf k},{\bf 0}}\nonumber\\
&&+\frac{ng^2}{V}\sum_{j}|A_j|^2\left|\frac{I_{\bf k}(t)-I_{\bf 0}(t)}{\varepsilon_{\bf k}}\right|^2,
\end{eqnarray}
with the impurity-depleted Bose condensate asymptotically given by
\begin{eqnarray}\label{}
\frac{\langle \Psi(t)|N_{\bf 0}|\Psi(t)\rangle}{N}=1+\frac{2|g|}{V}\sum_{j}|A_j|^2\left.\frac{\partial \Re I_{\bf k}(t)}{\partial \varepsilon_{\bf k}}\right|_{{\bf k}= {\bf 0}}.
\end{eqnarray}
The large-$|{\bf k}|$ tail of the momentum distribution (\ref{N_k}) fits the universal Tan's \cite{Tan_2} behavior $\langle \Psi(t)|N_{\bf k}|\Psi(t)\rangle \sim \mathcal{C}(t)/{\bf k}^4$, with the contact parameter $\mathcal{C}(t)\propto |I_{\bf 0}(t)|^2$ (see insert in Fig.~\ref{S(t)_fig}). The contact parameter's characteristic oscillatory zero-temperature equilibration behavior resembles cases of static Fermi \cite{Liu_2020} and Bose \cite{Drescher_2021} polarons without the spin-dependent interaction. The condensate depletion, controlled by the real part of $\ln f(t)$ (see Fig.~\ref{lnf(t)_fig}), is the oscillating function with increasing amplitude. We have also plotted in Fig.~\ref{moment_dis_fig}	
a function $\left|\frac{I_{\bf k}(t)-I_{\bf 0}(t)}{\varepsilon_{\bf k}}\right|^2$ (in dimensionless units) determining the time evolution of the momentum distribution of bosons in Eq.~\ref{N_k} after quench. The difference in behaviors for $|{\bf k}a|\ll 1$ and $|{\bf k}a|\ge 1$ can be understood by taking into account the long-time asymptotic expansion of the function $I_{\bf k}(t)$ from the Appendix. For $t|\epsilon_b|\gg 1$ and simultaneously $\varepsilon_{\bf k}t\gg 1$
\begin{eqnarray}
I_{\bf k}(t)\sim 2\theta(a)\frac{e^{-it\varepsilon_{\bf k}}-e^{it|\epsilon_b|}}{1+({\bf k}a)^2}-\frac{e^{-it\varepsilon_{\bf k}}}{1+({\bf k}a)^2}\nonumber\\
-\frac{(it|\epsilon_b|)^{-3/2}}{2\sqrt{\pi}({\bf k}a)^2}+O((it|\epsilon_b|)^{-5/2}),
\end{eqnarray}
if the second condition is violated, i.e. $\varepsilon_{\bf k}t\ll 1$, we find
\begin{eqnarray}
	I_{\bf k}(t)\sim 2\theta(a)[1-e^{it|\epsilon_b|}]-1
	+\frac{1}{\sqrt{\pi it|\epsilon_b|}}+\dots.
\end{eqnarray}
Thus, at initial times, the occupations of low-lying energy levels $\varepsilon_{\bf k}/|\epsilon_b|\ll 1$ oscillate (for positive $a$s) and increase up to times $\varepsilon_{\bf k}t\sim 1$ when the oscillation amplitudes start to decrease as a power-law function. A similar picture is obtained for the occupations of the higher energy levels $\varepsilon_{\bf k}/|\epsilon_b|\sim 1$, but magnitudes of amplitudes are smaller and start to decrease at much earlier times. 

\begin{figure}[h!]
	{\includegraphics
		[width=0.4
		\textwidth,clip,angle=-0]{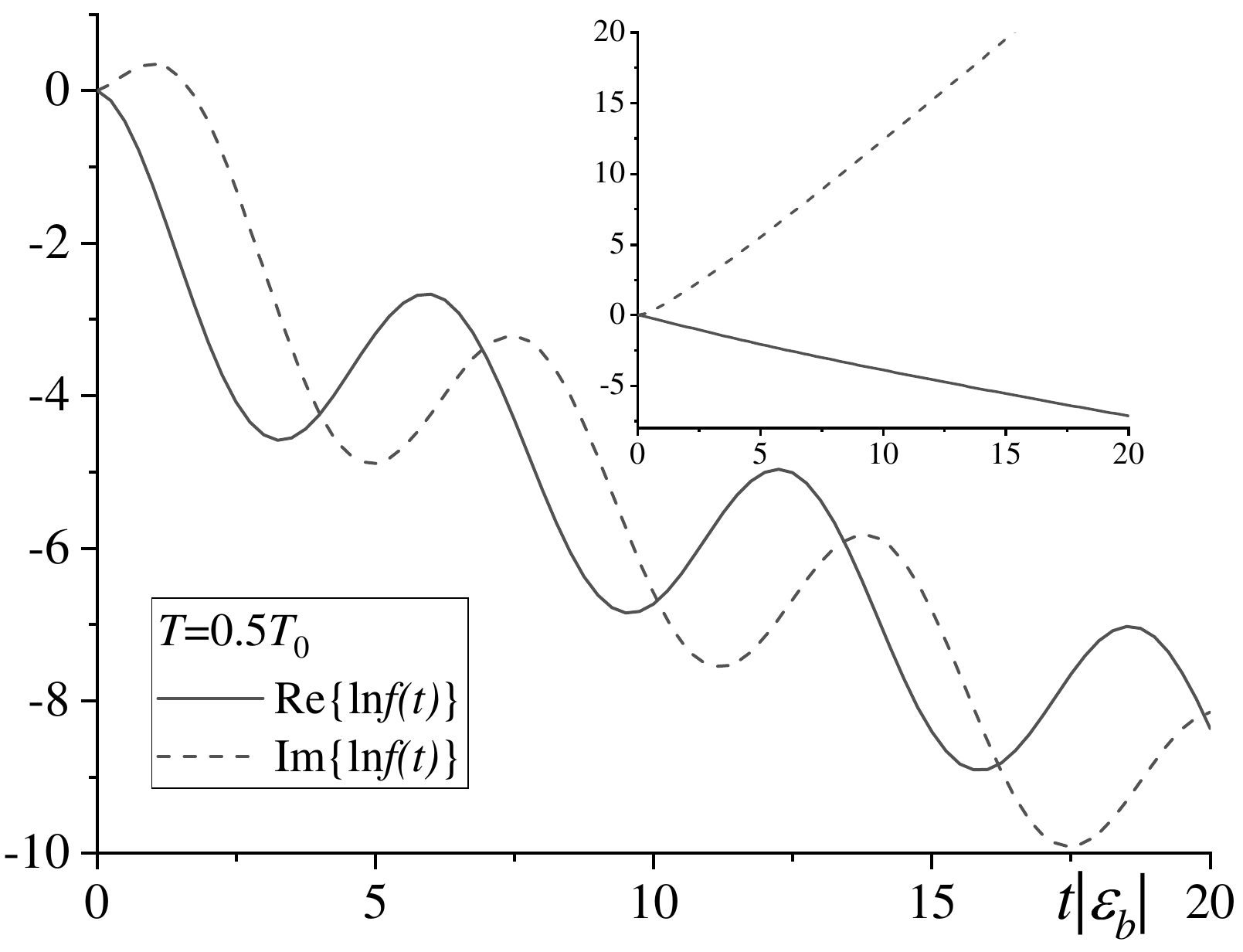}
	\includegraphics
	[width=0.4
	\textwidth,clip,angle=-0]{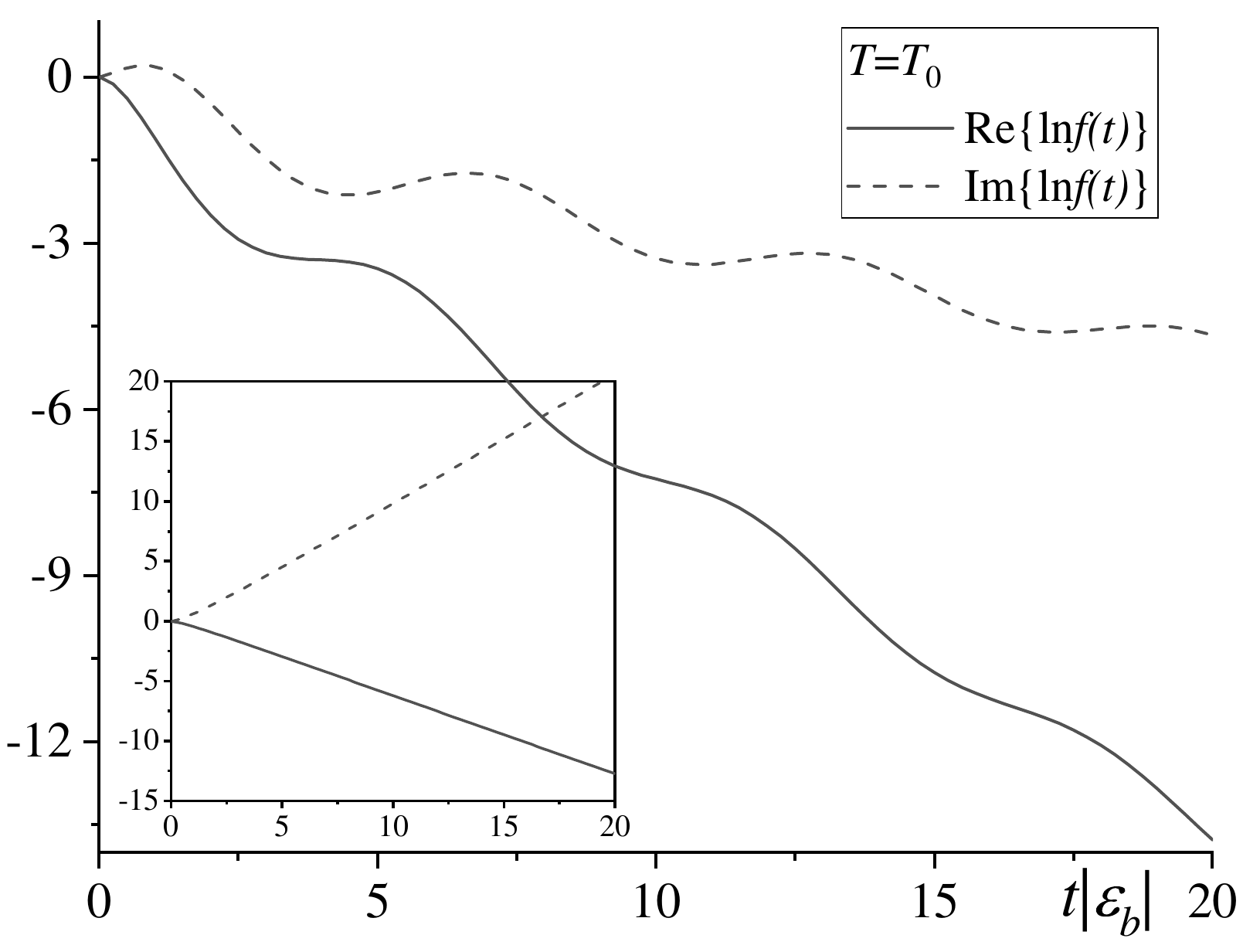}
\includegraphics
[width=0.4
\textwidth,clip,angle=-0]{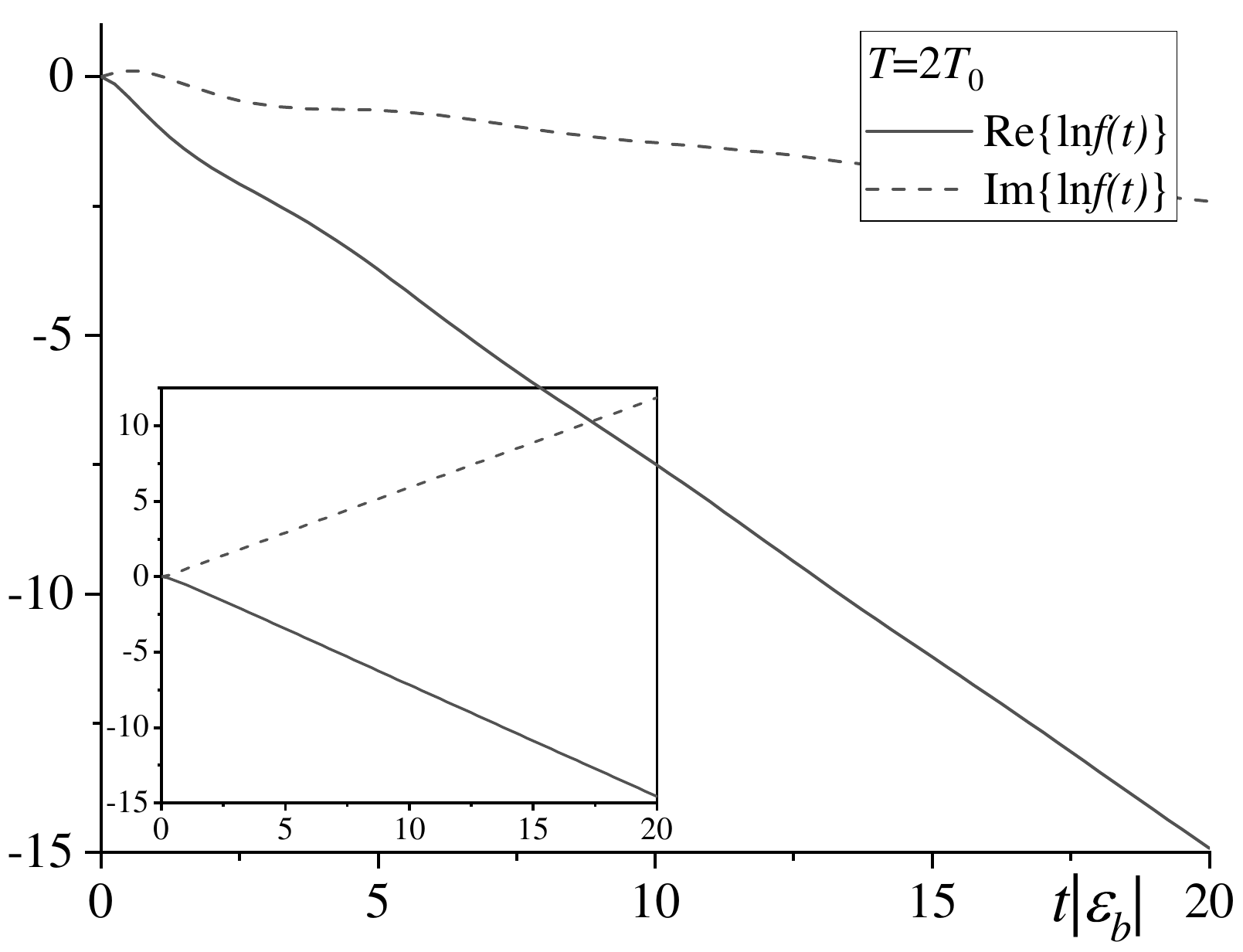}}
	\caption{Real (solid line) and imaginary (dashed line) parts of $\ln f_T(t)$ at finite temperatures (same notations as in Fig.~\ref{lnf(t)_fig}). Main panels show results for $a>0$, insets for $a<0$.}\label{lnf(t)_T_fig}
\end{figure}

At finite temperatures, we mainly focused on a study of the spin decoherence (see Fig.~\ref{lnf(t)_T_fig}).
There are two distinct regions where one should expect quite different behaviors: below and above the BEC-transition temperature $T_0$. This is because, at $T>T_0$, all correlations are exponentially suppressed by the chemical potential. As a consequence, the curves for $T=2T_0$ have small oscillations at early times and then almost linear decreases. Figuratively, the situation resembles that of the quantum field theory, where there is a contrast between interactions mediated by massless and massive gauge bosons. Temperatures below $T_0$ correspond to a massless case. Similar effects govern polarons' properties \cite{Mistakidis_2023,Grusdt_2025} and induced interaction between bipolarons \cite{Naidon_2018,Camacho-Guardian_2018,Panochko_2022,Jager_2022,Drescher_2023} in the dilute BECs. Figures~\ref{lnf(t)_T_fig} demonstrate apparent tendency: real part of $\ln f_T(t)$, which determines exponential decay of the off-diagonal matrix elements in (\ref{f_T}), increase faster in time at higher temperatures. An imaginary part slows down instead.

There is a distinct contrast between positive and negative couplings in all calculated observable quantities: there are no characteristic oscillations in the latter case. Regions where the derivative $\frac{\partial \Re\ln f_T(t)}{\partial t}>0$ are referred to \cite{Rivas_2014} as non-Markovian when the information flows from the environment to the system. Such an oscillating behavior is caused by the Bose statistics and the boson-impurity vacuum bound state, which is realized only for positive $g$s. In the case of non-interacting fermions or any other medium with non-zero compressibility, the two cases should not be that different. The latter is already apparent from results (see lower panel in Fig.~\ref{lnf(t)_T_fig}) for temperatures above the BEC transition point. It is important to preserve all terms in the double sum over wave-vectors in (\ref{Phi}) for a correct description of the boson-impurity bound state. A naive approximation that involves only condensate terms and neglects products of $b^{\dagger}_{\bf k}b_{\bf q}$ (with both ${\bf k}, {\bf q}\neq 0$) cannot capture the bound-state formation. 

\section{Summary}
We have analyzed an exactly solvable model of spin decoherence at zero and finite temperatures in a bath formed by an ideal Bose gas. The model suggests that bosons interact via a short-range potential with the infinitely heavy spin-$1/2$ impurity when its state is $|\uparrow\rangle$. It is demonstrated, however, that this simplified case covers (via unitary transformation) the most general boson-impurity interactions with spin-flipping. By solving the Heisenberg-like equations for bosonic creation and annihilation operators, we were able to calculate the time evolution of the wave function or density matrix of the initially non-interacting system. Then, tracing out bosonic degrees of freedom, one obtains the reduced density matrix of spin. Two distinct situations were considered: the zero-temperature evolution in the medium of bosons in the BEC state and the finite-temperature case with bosons in thermal equilibrium. We have found that both the entanglement properties and decoherence dynamics crucially depend on the sign of the boson-impurity coupling. A sudden increase in the interaction strength drives the bosons prepared initially in the Bose-condensed state to occupy excited energy levels. The calculated momentum distribution displays quite different time dependences for attractive and repulsive couplings.

\begin{center}
		{\bf Acknowledgements}
\end{center}
We thank Dr.~A.~Kuzmak for numerous discussions and comments regarding the manuscript. We are also grateful to Prof.~J.~Levinsen for his interest in the work and for providing the relevant references. The work of O.H. was supported by Project 2025.07/0326 (No.~0126U002943) from the National Research Foundation of Ukraine.

\section{Appendix}
Integral in the Mellin transform can be rewritten as follows:
\begin{eqnarray*}
C_{{\bf q},{\bf k}}(t)=\int_{-\infty}^{\infty}\frac{d\omega}{2\pi}e^{i\omega t}\Gamma_{{\bf q},{\bf k}}(i\omega+0_+),
\end{eqnarray*}
where a positive infinitesimal $0_+$ shifts poles from the real axis. In particular, this choice correctly reproduces $C_{{\bf q},{\bf k}}(t)$ without spin-boson interaction. Then, by using the spectral representation of the $T$-matrix
\begin{eqnarray*}
&&\mathcal{T}(i\omega+0_+)=\frac{2\theta(a)g|\epsilon_b|}{|\epsilon_b|-\omega+i0_+}\nonumber\\
&&-|g|\int_{0}^{\infty}\frac{d\nu}{\pi}\frac{\sqrt{\nu/|\epsilon_b|}}{1+\nu/|\epsilon_b|}\frac{1}{\nu+\omega-i0_+},
\end{eqnarray*}
[where $\theta(a)$ is the Heaviside theta function] one obtains
\begin{eqnarray*}
C_{{\bf q},{\bf k}}(t)=\delta_{{\bf q},{\bf k}}e^{-it\varepsilon_{\bf q}}+\frac{|g|}{V}\frac{I_{\bf k}(t)-I_{\bf q}(t)}{\varepsilon_{\bf k}-\varepsilon_{\bf q}}|\uparrow\rangle\langle \uparrow|,
\end{eqnarray*}
with the shorthand notations for the function
\begin{eqnarray*}
&&I_{\bf k}(t)=2\theta(a)|\epsilon_b|\frac{e^{-it\varepsilon_{\bf k}}-e^{it|\epsilon_b|}}{|\epsilon_b|+\varepsilon_{\bf k}}\nonumber\\
&&-\int_{0}^{\infty}\frac{d\nu}{\pi}\frac{\sqrt{\nu/|\epsilon_b|}}{1+\nu/|\epsilon_b|}\frac{e^{-it\varepsilon_{\bf k}}-e^{-it\nu}}{\nu-\varepsilon_{\bf k}}.
\end{eqnarray*}

\end{document}